\documentclass[12pt,epsf]{article}
\usepackage{graphicx}
\begin{document}

\def\a{\alpha}
\def\b{\beta}
\def\c{\varepsilon}
\def\d{\delta}
\def\e{\epsilon}
\def\f{\phi}
\def\g{\gamma}
\def\h{\theta}
\def\k{\kappa}
\def\l{\lambda}
\def\m{\mu}
\def\n{\nu}
\def\p{\psi}
\def\q{\partial}
\def\r{\rho}
\def\s{\sigma}
\def\t{\tau}
\def\u{\upsilon}
\def\v{\varphi}
\def\w{\omega}
\def\x{\xi}
\def\y{\eta}
\def\z{\zeta}
\def\D{\Delta}
\def\G{\Gamma}
\def\H{\Theta}
\def\L{\Lambda}
\def\F{\Phi}
\def\P{\Psi}
\def\S{\Sigma}

\def\o{\over}
\newcommand{\bear}{\begin{array}}  \newcommand{\eear}{\end{array}}
\newcommand{\bea}{\begin{eqnarray}}  \newcommand{\eea}{\end{eqnarray}}
\newcommand{\beq}{\begin{equation}}  \newcommand{\eeq}{\end{equation}}
\newcommand{\bef}{\begin{figure}}  \newcommand{\eef}{\end{figure}}
\newcommand{\bec}{\begin{center}}  \newcommand{\eec}{\end{center}}
\newcommand{\non}{\nonumber}  \newcommand{\eqn}[1]{\beq {#1}\eeq}
\newcommand{\lmk}{\left(}  \newcommand{\rmk}{\right)}
\newcommand{\lkk}{\left[}  \newcommand{\rkk}{\right]}
\newcommand{\lhk}{\left \{ }  \newcommand{\rhk}{\right \} }
\newcommand{\del}{\partial}  \newcommand{\abs}[1]{\vert{#1}\vert}
\newcommand{\vect}[1]{\mbox{\boldmath${#1}$}}
\newcommand{\bib}{\bibitem} \newcommand{\new}{\newblock}
\newcommand{\la}{\left\langle} \newcommand{\ra}{\right\rangle}
\newcommand{\bfx}{{\bf x}} \newcommand{\bfk}{{\bf k}}
\newcommand{\gtilde} {~ \raisebox{-1ex}{$\stackrel{\textstyle >}{\sim}$} ~} 
\newcommand{\ltilde} {~ \raisebox{-1ex}{$\stackrel{\textstyle <}{\sim}$} ~}
\newcommand{\gtrsim}{ \mathop{}_{\textstyle \sim}^{\textstyle >} }
\newcommand{\lesssim}{ \mathop{}_{\textstyle \sim}^{\textstyle <} }
\newcommand{\ds}{\displaystyle}
\newcommand{\gsim}{ \mathop{}_{\textstyle \sim}^{\textstyle >} }
\newcommand{\lsim}{ \mathop{}_{\textstyle \sim}^{\textstyle <} }
\newcommand{\vev}[1]{ \left\langle {#1} \right\rangle }
\newcommand{\bra}[1]{ \langle {#1} | }
\newcommand{\ket}[1]{ | {#1} \rangle }
\newcommand{\EV}{ {\rm eV} }
\newcommand{\KEV}{ {\rm keV} }
\newcommand{\MEV}{ {\rm MeV} }
\newcommand{\GEV}{ {\rm GeV} }
\newcommand{\TEV}{ {\rm TeV} }
\def\diag{\mathop{\rm diag}\nolimits}
\def\Spin{\mathop{\rm Spin}}
\def\SO{\mathop{\rm SO}}
\def\O{\mathop{\rm O}}
\def\SU{\mathop{\rm SU}}
\def\U{\mathop{\rm U}}
\def\Sp{\mathop{\rm Sp}}
\def\SL{\mathop{\rm SL}}
\def\tr{\mathop{\rm tr}}

\def\IJMP{Int.~J.~Mod.~Phys. }
\def\MPL{Mod.~Phys.~Lett. }
\def\NP{Nucl.~Phys. }
\def\PL{Phys.~Lett. }
\def\PR{Phys.~Rev. }
\def\PRL{Phys.~Rev.~Lett. }
\def\PTP{Prog.~Theor.~Phys. }
\def\ZP{Z.~Phys. }

\def\lrf#1#2{ \left(\frac{#1}{#2}\right)}
\def\lrfp#1#2#3{ \left(\frac{#1}{#2}\right)^{#3}}


\baselineskip 0.7cm

\begin{titlepage}

\begin{flushright}
DESY 07-019
\end{flushright}

\vskip 1.35cm
\begin{center}
{\large \bf
   Retrofitted Gravity Mediation without the Gravitino-overproduction Problem
}
\vskip 1.2cm
Motoi Endo${}^{1}$, Fuminobu Takahashi${}^{1}$ 
and T. T. Yanagida${}^{2,3}$
\vskip 0.4cm

${}^1${\it   Deutsches Elektronen Synchrotron DESY, \\
Notkestrasse 85,  22607 Hamburg, Germany}\\
${}^2${\it  Department of Physics, University of Tokyo,\\
     Tokyo 113-0033, Japan}\\
${}^3${\it Research Center for the Early Universe, University of Tokyo,\\
     Tokyo 113-0033, Japan}

\vskip 1.5cm

\abstract{ We propose a retrofitted gravity mediation model which
alleviates the gravitino overproduction from decays of an inflaton and a
supersymmetry breaking field. In the model, we introduce an
approximate $U(1)$ symmetry under which the supersymmetry breaking
field is charged, although it is broken by a mass term of messenger
fields to generate gaugino masses of order the weak scale.  In a
low-scale inflation model, we find regions in which the gravitino
overproduction problem is avoided.  }
\end{center}
\end{titlepage}

\setcounter{page}{2}

\section{Introduction}

Supersymmetry (SUSY) is one of the most plausible candidates for a
theory beyond the standard model. Since we have not observed any
supersymmetric partners of the standard-model particles yet, SUSY must
be broken in the vacuum.  The central issue is how to mediate the SUSY
breaking effect to the visible sector. Among many scenarios proposed
so far, gravity mediated SUSY breaking models have been
thoroughly and continuously studied~\cite{Nilles:1983ge}.  The gravity
mediation has both good and bad points. It can naturally generate the
$\mu$ term of the desired magnitude~\cite{GM,IKYY} and may also
explain the dark matter (DM) abundance by the lightest supersymmetric
particle (LSP), while it is plagued with the SUSY flavor and CP
problems.  In spite of its potential problems, the gravity mediation
has attracted considerable attention as the simplest mediation
mechanism of the SUSY breaking.

In recent articles~\cite{KTY} it has been pointed out that there is a
new gravitino overproduction problem in supergravity (SUGRA). That is,
many gravitinos are produced directly in the decay of the inflaton if
the inflaton has a non-vanishing vacuum-expectation value (VEV). The
detailed analyses have shown that most of the parameter space in the
gravity-mediation model of SUSY breaking is excluded by this direct
gravitino-production process together with the gravitino production by
particle scatterings in thermal bath.  A crucial point here is that
the gravity mediation has a singlet field $Z$ responsible for the SUSY
breaking, which mixes with the inflaton field~\cite{Endo:2006tf}.
Because of this mixing the inflaton decays into a pair of gravitinos.

Furthermore, the gravity-mediation model suffers from the Polonyi
problem~\cite{Polonyi}.  Since the $Z$ field should be completely
neutral under any symmetries to generate the gaugino
masses~\cite{Nilles:1983ge}, the origin of the $Z$ field has no
special meaning and hence the value of $Z$ during inflation is
generically different from the minimum in the true vacuum. Therefore,
the potential energy of the $Z$ will dominate the universe after the
inflation and its decay destroys the light nuclei produced by big bang
nucleosynthesis (BBN), or if the mass of $Z$ is larger than $2\times
m_{3/2}$ it decays into a pair of gravitinos. (Here, $m_{3/2}$ is the
gravitino mass.) The succeeding decay of the gravitino destroys again
the light nuclei and ruins the success of BBN.  It has been recently
stressed that this problem is not solved even in the case of dynamical
SUSY breaking~\cite{Ibe:2006am}.

All above problems are originated from an assumption that the $Z$
field responsible for the SUSY breaking is completely neutral.  To
avoid the problems while keeping the merits of the gravity mediation
stated at the beginning of this section~\footnote{
The purpose of this paper is not to solve the SUSY flavor and CP problems,
but to alleviate the cosmological problems that the original gravity mediation has.
}, we propose a
gravity-mediation model with a $Z$ field charged under some
symmetry. In this model the gaugino masses vanish at the tree level,
since couplings of $Z$ to the kinetic functions of the gauge
multiplets are forbidden by the symmetry.  So we introduce a pair of
messengers whose mass term breaks the symmetry, to generate the
gaugino masses. The one-loop diagrams of the messengers, in fact,
induce the gaugino masses picking up the symmetry-breaking mass term
of the messengers.  We show that the present gravity-mediation model
indeed relaxes the gravitino-overproduction problem mentioned above if
the inflation scale $H_{\rm inf}$ is sufficiently low as $H_{\rm inf}
\lsim {\rm a~few}\times 10^6$ GeV.

The paper is organized as follows. In Sec.~\ref{sec:2} we describe the
retrofitted gravity mediation model. We discuss the cosmology of our
model in Sec.~\ref{sec:3}, particularly focusing on the gravitino
production from both the SUSY breaking field and the inflaton. The
last section is devoted to conclusions.

\section{A retrofitted gravity-mediation model}
\label{sec:2}

The model is based on a dynamical SUSY-breaking model proposed
in~\cite{IYIT}, which assumes an $SP(1)$ gauge theory with 4 chiral
superfields $Q^i~ (i=1-4)$ in the $SP(1)$ doublet representation,
where the gauge index is omitted.  Without a superpotential this
theory possesses a flavor $SU(4)_F$ symmetry. We assume, for
simplicity, that the flavor symmetry is explicitly broken down to an
$SP(2)_F$ by a superpotential.  Thus we introduce 5 gauge singlet
superfields $Z_a ~(a=1-5)$ and assume the tree-level superpotential
\begin{equation}
      W_0 \;=\; \lambda ' Z_a(QQ)^a,
\end{equation}
where $(QQ)^a$ denotes a flavor 5-plet of the $SP(2)_F$ given by a
suitable combination of $SP(1)$ gauge invariants $Q^{i}Q^j$. Together
with the effective superpotential induced by the strong $SP(1)$ gauge
interactions,
\begin{equation}
      W_{\rm dyn} \;=\; X({\rm Pf}(Q^iQ^j)-\Lambda^4)
\label{eq:w_dyn},
\end{equation}
the superpotential $W_{\rm dyn}$ implies that the $SP(2)_F$ singlet
$(QQ)=\frac{1}{2}(Q^1Q^3+Q^2Q^4)$ condensates and we find
\begin{equation}
      \langle (QQ)\rangle \;=\; \Lambda^2.
\end{equation}

We further introduce an $SP(2)_F$ singlet superfield $Z$ and consider a
tree level superpotential,
\begin{equation}
\label{eq:zqq}
      W \;=\; W_0 + \lambda Z(QQ).
\end{equation}
For the coupling $\lambda \le O(1)$, we find the vacuum,
\begin{equation}
      \langle (QQ)\rangle \; =\; \Lambda^2,~~~~\langle (QQ)^a\rangle \simeq 0. \label{eq:vauum}
\end{equation}
After integrating the massive modes we have the low-energy effective
superpotential
\begin{equation}
      W_{\rm eff} \;\simeq\; \lambda \Lambda^2Z,
\label{eq:lowE}      
\end{equation}
which yields a dynamical SUSY breaking~\cite{IYIT},
\begin{equation}
       F_Z \;\simeq\; \lambda \Lambda^2.
\label{eq:f_z}
\end{equation}

Notice that the tree-level superpotential Eq.~(\ref{eq:zqq}) possesses
a global $U(1)$ symmetry at the classical level, under which the $Z$
fields and $Q^i$ transform as
\begin{equation}
    Z\rightarrow e^{-i\delta}Z,~~~~~Q^i\rightarrow e^{+\frac{i}{2}\delta} Q^i.\label{eq:u(1)}
\end{equation}
We use this global $U(1)$ to avoid the gravitino-overproduction and
the Polonyi problems as explained in the introduction
\footnote{A discrete $Z_2$ is sufficient for our purpose where the
$Z,~QQ$ and $\Psi{\bar \Psi}$ have odd parity of the $Z_2$.}, although
it is broken by $SP(1)$ instanton effects at the quantum level (see
also Eq.~(\ref{eq:w_dyn})).

In SUGRA the gravitino acquires a SUSY-breaking mass $m_{3/2}$ from 
Eq.~(\ref{eq:f_z}) as~\cite{Nilles:1983ge}
\begin{equation}
      m_{3/2} \simeq \frac{F_Z}{\sqrt{3}}.
      \label{eq:m_3/2}
\end{equation}
Here and in what follows, we adopt the Planck unit, $M_P =1$ unless
otherwise stated, where $M_P\simeq 2.4\times 10^{18}$GeV is the
reduced Planck scale.  For a generic K\"ahler potential squarks,
sleptons and Higgs bosons acquire the SUSY-breaking soft masses of
$O(m_{3/2})$. However, the gauginos in the SUSY standard model (SSM)
remain massless~\cite{Banks:1993en,Dine:1992yw}, since the $Z$ does
not have couplings to the gauge kinetic functions. In fact, the
interaction,
\begin{equation}
      \int d^2\theta ZW^a_\alpha W^a_\alpha ,
\end{equation}
is forbidden by the $U(1)$ symmetry in Eq.~(\ref{eq:u(1)}), where
$W^a_\alpha$ are chiral superfields for gauge multiplets.  Therefore,
we need a breaking term of the global $U(1)$ symmetry to generate
gaugino masses. Otherwise, the dominant contribution to the gaugino
masses comes only from the scale-invariance anomalies at the quantum
level~\cite{anomaly}, which may be of order $10^{-2}\times m_{3/2}$.
For $m_{3/2}=100~{\rm GeV}-10$ TeV we have the gaugino masses of order
$1-100$ GeV which is excluded already by experiments.  Thus, to have
larger gaugino masses we introduce a pair of messengers $\Psi$ and
${\bar \Psi}$ whose mass term breaks the global $U(1)$, assuming that
they transform as ${\bf 5}$ and ${\bf 5^*}$ of $SU(5)_{\rm GUT}$,
respectively.

\begin{table}[t]
\begin{center}
\begin{tabular}{|l|c|c|c|c|}
\hline \hline
 & $Z$ & $QQ$ & $\Psi{\bar \Psi}$ & M \\ \hline
$U(1)_R$ & $0$ & $+2$ & $+2$ & $0$ \\ \hline
$U(1)$& $+1$ & $-1$ & $-1$ & $+1$ \\ 
\hline \hline
\end{tabular}
\caption{The charges of $U(1)_R$ and $U(1)$.}
\end{center}
\end{table}

Then, the $Z$ field has a superpotential with the messenger fields as
\begin{equation}
      W_{\rm messenger}= kZ\Psi{\bar \Psi} + M\Psi{\bar \Psi}.
\label{eq:messenger}      
\end{equation}
The $U(1)$ and $U(1)_R$ charges for relevant superfields are given in
Table 1. We see that the messenger mass term $M$ breaks the global
$U(1)$ symmetry. Here and in what follows, we assume that, at the
breaking scale $M$, the effect of the $U(1)$ breaking appears only in
the mass term of $\Psi$ and $\bar \Psi$~\footnote{ If one allows any
$U(1)$-breaking operators suppressed by powers of $M$, there are such
dangerous operators as $K \simeq M^\dag Z |\phi|^2+{\rm h.c.}$ which
induce the severe gravitino overproduction, where $\phi$ denotes the
inflaton field. }.  The integration of the messengers give rise to the
gaugino masses as~\cite{GMSB}~\footnote{
Although the scalar trilinear couplings are suppressed at the dynamical
scale in this model, it is possible to induce sizable contributions by
introducing the Yukawa interactions, $Y_S S H \bar\Psi + Y'_S S \bar H
\Psi$, with the SM singlet $S$, assigning the $U(1)$ and $U(1)_R$
charges for $S$ and $H, \bar H$
properly~\cite{Giudice:1997ni,Yoshioka}.  }
\begin{equation}
      m_i \simeq \frac{\alpha_i}{4\pi} \frac{kF_Z}{M} ~~~~~~~{\rm for}~~i=1,2,3.
      \label{eq:m_i}
\end{equation}
Here, $m_{1,2,3}$ and $\alpha_{1,2,3}$ are the gaugino masses and the
gauge coupling constants for $U(1), SU(2)$ and $SU(3)$ in the SSM. We
have used the $SU(5)_{\rm GUT}$ normalization for the $U(1)$ gauge
coupling constant.  For $m_3\simeq 1$ TeV we have
\begin{equation}
      \frac{kF_Z}{M} \simeq 10^5~{\rm GeV}.\label{eq:kf/m}
\end{equation}

Notice that the global $U(1)$ and $U(1)_R$ charges for the operator
$QQ$ are the same as the $\Psi{\bar \Psi}$ and hence the dynamical
quarks may naturally have a mass term $M'QQ$ with $M'\simeq O(M)$. In
the text, we have redefined the $Z$ field by a shift , $Z\rightarrow
Z-M'$, to absorb the mass term for the dynamical quarks $Q$.  However,
this shift of the field induces, for instance, a linear term of $Z$ in
the K\"ahler potential (see Eq.~(\ref{eq:linear})). In the following,
we adopt the origin of $Z$ as that obtained after the shift.

We should mention here that the Giudice-Masiero(GM)
mechanism~\cite{GM} for generating a SUSY-invariant mass term (called
the $\mu$ term), $\mu H{\bar H}$, for Higgs doublets does not work,
provided that the Higgs multiplets, $H$ and ${\bar H}$ are neutral of
the global $U(1)$ symmetry~\footnote{
If one assumes the $U(1)$-charge $+1$ for $H{\bar H}$, the K\"ahler
coupling $Z^\dagger H{\bar H}$ is allowed and the GM mechanism works.
}.
However, they receive the $\mu$ term though the following 
superpotential~\cite{IKYY},
\begin{equation}
      W= C(1+hH{\bar H}),
\end{equation}
where $C=m_{3/2}$ is the constant introduced to cancel the
vacuum-energy density $|F_Z|^2$ for the SUSY-breaking. Here, we have
assumed that $H{\bar H}$ carries a vanishing $U(1)_R$ charge. Then, we
find the $\mu$ parameter for the Higgs mass as
\begin{equation}
       \mu = m_{3/2}\times h.
\end{equation}

In the following discussion we restrict ourselves to the parameter
region of so-called gravity mediation, that is, $m_{3/2}\simeq 100
{\rm GeV}-10$ TeV and the gluino mass $m_{3}\simeq 1$ TeV.  This
implies from Eqs.~(\ref{eq:m_3/2}), (\ref{eq:m_i}) and (\ref{eq:kf/m})
\begin{equation}
       \sqrt{F_Z}\;=\;
       \sqrt{\lambda }\Lambda
       \simeq 2\times 10^{10}{\rm GeV}-2\times 10^{11} {\rm GeV},
       \label{eq:sf_z}
\end{equation}
and
\begin{equation}
       \frac{k}{M}\simeq 3\times 10^{-(16-18)} {\rm GeV}^{-1}.
\end{equation}
We see that the SUSY-preserving vacuum, $\langle Z\rangle = -M/k
\simeq 4\times 10^{15-17}$ GeV with $\langle \Psi{\bar \Psi} \rangle
\simeq -\lambda\Lambda^2/k$, is far from the dynamical scale $|Z|
\simeq \Lambda$ in Eq.~(\ref{eq:sf_z}), and hence the SUSY-breaking
vacuum, $|\langle Z\rangle | \lsim \Lambda$ and $F_Z\simeq
\lambda\Lambda^2$, is practically stable. Indeed, the messenger fields
are not tachyonic at the origin $Z = 0$ and the tunneling rate is also
suppressed as long as $M \gg \sqrt{k \lambda} \Lambda$, which is
satisfied unless $k$ is extremely small~\cite{metastable}.

\section{Cosmology}
\label{sec:3}
\subsection{Polonyi problem}
\begin{figure}[t!]
\begin{center}
\includegraphics[width=7cm]{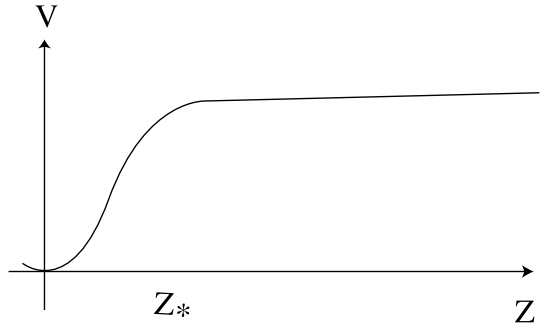}
\caption{Schematic potential of the scalar component of the $Z$ field.}
\label{fig:potential}
\end{center}
\end{figure}

Let us first discuss the Polonyi problem in the present model.  We
assume that the $SP(1)$ hadrons have masses of order $4\pi \Lambda$
and hence above the scale $|Z| > |Z_*| \equiv 4\pi \Lambda/\lambda$,
the Polonyi field $Z$ does not receive effects from the $SP(1)$ strong
interactions.  The potential of $Z$ is therefore very flat above
$|Z_*|$.  On the other hand, for $|Z| < |Z_*|$ the $Z$ acquires a
larger SUSY-breaking soft mass from the loop diagrams of the $SP(1)$
hadrons, and it is given by
\begin{equation}
       m_{Z} \simeq \frac{\eta}{16\pi^2}\lambda^{3} \Lambda,
       \label{eq:massZ}
\end{equation}
where $\eta$ is a numerical coefficient which is 
expected to be order unity.
The potential for $Z$ can be approximated by~\cite{Arkani-Hamed:1997ut}
\beq
V_L(Z) \simeq \left\{
\bear{cc}
m_Z^2 |Z|^2 & {\rm~~~~~for~}|Z| < |Z_*| \\
&\\
\xi \, m_{3/2}^2 & {\rm~~~~~for~}|Z| >  |Z_*|
\eear
\right.,
\eeq
where we have set the cosmological constant to zero at the origin, and
$\xi \lesssim O(1)$ is a real constant. Note that, for $|Z| > |Z_*|$,
the potential is nonzero due to the $SP(1)$ gaugino condensation, and
the curvature of the potential is given by~\cite{Ibe:2006am}
\beq
V_L''(Z) \simeq - \frac{3 \lambda^2}{4 \pi^2}\frac{m_{3/2}^2 }{|Z|^2} 
+ O(m_{3/2}^2),
\label{eq:v''}
\eeq
where the first term comes from the perturbative wavefunction
renormalization of $Z$, while the second term represents the
contribution from the gravity mediation.  For $\lambda \simeq O(1)$
and $|Z| \ll 1$, the first term dominates over the second term, and so,
we will focus on the first term from here on.  A
schematic potential for the $Z$ boson is shown in
Fig.~\ref{fig:potential}.

If the global $U(1)$ symmetry discussed in the previous section is
exact, the K\"ahler potential for the $Z$ is a function of $Z^\dagger
Z$ and the origin of the $Z$ is most likely the potential minimum not
only at the present, but also during the inflation.  However, since we
introduce a mass parameter $M$ for the symmetry breaking, it is
natural to consider the K\"ahler potential contains breaking
terms. Thus we consider the K\"ahler potential for the $Z$
as~\footnote{ Even if the K\"ahler potential does not contain such
symmetry breaking term from the beginning, it is radiatively induced
as $K \sim \frac{1}{16 \pi^2}k M^\dag Z + {\rm h.c.}$.  }
\begin{equation}
       K= (\delta M^\dagger Z + {\rm h.c.})+ ZZ^\dagger +  \cdots,
       \label{eq:linear}
\end{equation}
where $\delta$ is a constant of $O(1)$.

We see that the minimum of the $Z$ potential shifts from that at the
vacuum during the inflation. The effective potential for $Z$ during
inflation is given by~\footnote{ It should be noted that, during
inflation, $V_L(Z)$ is absent for a high-scale inflation model with
$H_{\rm inf} > 4 \pi \Lambda$. However, this does not change our
argument.}
\bea
V(Z) &\simeq& e^{K} (3 H_{\rm inf}^2  )+ V_L(Z),\non\\
	  &\simeq& 3 H_{\rm inf}^2 \left( |Z|^2 + \delta M^\dag Z + \delta M Z^\dag  + \cdots\right) 
	    	    + V_L(Z).
\eea
If the Hubble parameter during the inflation satisfies $H_{\rm inf} >
m_{Z}$, the shift is therefore~\footnote{ Such a large deviation from
the origin is dangerous not only because it induces the cosmological
disaster discussed below, but also because the $Z$ field may roll down
to the supersymmetric vacuum after inflation.  }
\begin{equation}
       |\Delta Z|\simeq |\delta M| \gg Z_*.
\end{equation}
After inflation, the $Z$ field starts to oscillate with an initial
amplitude $|\Delta Z|$, when the Hubble parameter becomes comparable
to the curvature of the potential at $Z \simeq |\Delta Z|$. Since the
potential has an approximate $U(1)$ symmetry and is almost flat for
$|Z| > |Z_*|$, the $Z$ field experiences spatial instabilities and
soon deforms into $Q$-balls~\cite{Coleman:1985ki} with a typical
charge~\cite{Kasuya:1999wu},
\bea
Q &\simeq&
6 \times 10^{-4} \frac{|\Delta Z|^4}{|V''_L(\Delta Z)|^2}
,\non\\&\simeq& 
2 \times 10^{21} \,\frac{\delta^4 k^4}{\lambda^2} \lrfp{\alpha_3}{0.1}{4} 
 \lrfp{m_3}{\rm TeV}{-4}\lrfp{m_{3/2}}{{\rm TeV}}{2},
\eea
where we have used (\ref{eq:m_i}) in the second equality.  After the
$Q$-balls are formed, the energy density of $Z$ decreases as $a^{-3}$
($a$ is the scale factor) like a non-relativistic matter.  Since the
lifetime of the $Q$-ball is rather long as shown below, the $Z$ field
will dominate the energy of the universe before the decay.

How does the $Z$ field in the form of the $Q$-balls decay? The $Z$
field decays into a pair of the gravitinos as well as the SM fields.
Since the $Z$ decays only through (effective) higher dimensional
operators, the decay rate of the $Q$-balls is unlikely to saturate the
geometrical upper bound~\cite{Cohen:1986ct}.  For the moment we assume
this is not the case.  Then $Z$ decays in the entire volume of the
$Q$-balls.  Inside the $Q$-balls, the $Z$ field rotates in its
internal space with a constant angular velocity, and the $Q$-ball
solution is of the form:
\beq
Z(r,t) \;=\; Z(r)e^{i \omega t}.
\label{Qball}
\eeq
Here $\omega \simeq |V''(\Delta Z)|^{1/2}$ is the angular velocity 
and $Z(r)$ represents the radial profile of the $Q$-ball with 
a radial coordinate $r$, ranging from $r = 0$ to $r = R_Q$, 
where $R_Q$ is the radius of the $Q$-ball. The field value at the center
$r=0$, $Z(0)$, is roughly equal to $\Delta Z$. For later use, let
us express $\omega$ in terms of the gluino mass as
\beq
\omega \;\simeq\; \frac{2}{\alpha_3} \lrf{\lambda}{\delta k} m_3,
\label{eq:ome}
\eeq
where we have used (\ref{eq:m_i}) and (\ref{eq:v''}).  The
classical-field configuration (\ref{Qball}) is interpreted as a
condensate of the $Z$-particles with energy $\omega$ per unit quanta
and with a macroscopic number density $\sim \omega |\Delta Z|^2$.
Therefore one can use the perturbative decay rate by substituting
$\omega$ for the mass of $Z$~\footnote{ Since the finite size of the
$Q$-ball implies a finite momenta, $p \lesssim R_Q^{-1} \sim \omega$,
for a quanta in the condensate, numerical factors may change by $O(1)$
in the following decay rates. }.

Since the $Z$ field has a nonzero $F$-term $|F_Z| \sim m_{3/2}$ even
for $|Z| > |Z_*|$ due to the gaugino condensation, it decays into a
pair of the gravitinos. The decay rate is given by
\bea 
\Gamma_Z(Z \rightarrow 2 \psi_{3/2}) &\simeq& \frac{1}{96 \pi}
\frac{\omega^5}{m_{3/2}^2 },\non\\ 
&\simeq & 1
\times 10^{-24} {\rm\,GeV} \lrfp{\lambda}{\delta k}{5}
\lrfp{\alpha_3}{0.1}{-5} \lrfp{m_3}{\rm TeV}{5}\lrfp{m_{3/2}}{{\rm
TeV}}{-2},
 \label{eq:gammaz}
\eea
where $m_3$ and $m_{3/2}$ are evaluated at the weak scale.  In the
goldstino picture, the above decay into the gravitinos is induced by
the four point interaction $|Z|^4$ in the K\"ahler potential, which
arises from the wavefunction renormalization of $Z$.

In addition, $Z$ interacts with the SM gauge sector via the
messenger loops. Since $\omega \ll M$, we can write an effective
coupling as
\beq
\mathcal{L} \;=\; - \int d^2\theta \frac{\alpha_i}{8\pi} 
\frac{k Z}{M + kZ} W^{(i)}_\alpha W^{(i)}_\alpha + {\rm h.c.},
\eeq
where $Z$ is the superfield. 
This interaction is expanded as
\begin{eqnarray}
    \mathcal{L}\; \simeq\;
    -\frac{\alpha_i}{4\pi} \frac{k}{M} Z
    \bigg[ 
    - \frac{1}{4} F_{\mu\nu}^{(i)} F^{(i) \mu\nu} 
    + \frac{i}{8} \epsilon^{\mu\nu\rho\sigma} 
      F_{\mu\nu}^{(i)} F_{\rho\sigma}^{(i)} 
    - \frac{k F_Z}{ M} 
      \bar{\lambda}^{(i)} {\mathcal P}_L \lambda^{(i)} 
    \bigg] + {\rm h.c.},
\end{eqnarray}
where we neglected terms with higher orders of $k\langle Z \rangle/M$. 
Then the decay rate into the $SU(3)_C$ gluons is
\bea 
\Gamma_Z(Z \rightarrow 2 g) 
&\simeq & 
8\times \frac{k^2}{8 \pi} \lrfp{\alpha_3}{8 \pi}{2} \frac{\omega^3}{M^2},\non\\
&\simeq& 
4\times 10^{-26} {\rm GeV}
\lrfp{\lambda}{\delta k}{3} \lrfp{\alpha_3}{0.1}{-3} \lrfp{m_3}{\rm TeV}{5}
\lrfp{m_{3/2}}{{\rm TeV}}{-2},
 \label{eq:gammag}
\eea
while that into the gluinos is
\bea \Gamma_Z(Z \rightarrow 2 \tilde{g}) 
&\simeq & 
8\times \frac{1}{8 \pi} \lrfp{\alpha_3 k^2 F_Z}{4 \pi M^2}{2} \omega,\non\\
&\simeq& 
6\times 10^{-24} {\rm GeV}
\lrf{\lambda}{\delta k} \lrfp{\alpha_3}{0.1}{-3} \lrfp{m_3}{\rm TeV}{5}
\lrfp{m_{3/2}}{{\rm TeV}}{-2},
 \label{eq:gammagluino}
\eea
where we have assumed $\omega$ is much larger than $2m_3$.

The $Z$ also decays into the SM matters via the top Yukawa coupling,
since it has a linear term in the K\"ahler
potential~\cite{Endo:2006qk}.  However, the rate is smaller than that
into the gauge bosons. The decay via anomalies in
SUGRA~\cite{Endo:2007ih} is also suppressed by a loop factor, so we
neglect them here.

One can check that the decay rates (\ref{eq:gammaz}) and
(\ref{eq:gammag}) (or (\ref{eq:gammagluino}) ) are much smaller than
the geometrical upper bound on the $Q$-ball decay
rate:~\cite{Cohen:1986ct}
\bea
\Gamma_Q &=& \frac{1}{Q} \left|\frac{dQ}{dt} \right|,\non\\
                      &\simeq& 8 \times 10^{-20} {\rm \, GeV}  
                \lrf{\lambda^3}{\delta^5 k^5}      \lrfp{\alpha_3}{0.1}{-5} 
			 \lrfp{m_3}{\rm TeV}{5}\lrfp{m_{3/2}}{{\rm TeV}}{-2},
\label{Q}			 
\eea
The upper bound can be thought of as a dissipation rate of
relativistic decay products. If the decay products are fermions
(e.g. the gravitinos and gauginos in our case), and if the
perturbative decay rates obtained in a way described above exceed the
bound, the decay processes inside the $Q$-ball would be suppressed by
the Pauli blocking since the decayed products would fill the phase
space. Then, the dissipation rate would determine the decay rate of
the $Q$-ball. In our case, however, since the decay proceeds only
through the higher dimensional operators and the perturbative decay
rates are so small, such suppression is absent.  On the other hand, if
the decay products are bosons (e.g. the gluons in our case), there is
no Pauli blocking inside the $Q$-ball. However, if the bosons acquire
a large mass due to interactions with the scalar field that form the
$Q$-ball, the decay inside the $Q$-ball might be kinematically
blocked. Then the decay rate of the $Q$-ball would become again
$\Gamma_Q$.  In the present case, however, since the $Z$ field is
singlet under the SM gauge symmetry, the gluons are massless inside
the $Q$-balls. Thus the decay rate of the $Q$-ball is given by the
perturbative decay rate.

It is illustrative to take the ratios of  the above decay rates:
\bea
\frac{\Gamma_Z(Z \rightarrow 2 \psi_{3/2})}{ \Gamma_Z(Z \rightarrow 2 g) }
&\simeq& \frac{1}{2 \alpha_3^2} \lrfp{\lambda}{k \delta}{2} ,\\
\frac{\Gamma(Z \rightarrow 2 \tilde g)}{\Gamma(Z \rightarrow 2 g)}
&\simeq& 16\pi^2 \lrfp{\lambda}{k \delta}{-2}.
\eea
Therefore, for $(\lambda/k\delta) \gg 1$, the dominant decay mode is
that into the gravitinos, while the gluino production dominates over
the others for $0.1 \lesssim (\lambda/k\delta) \lesssim O(1)$.  For
$(\lambda/k\delta) \lesssim 0.1$, the decay into the gluinos is
kinematically forbidden (see (\ref{eq:ome})), and the gluon production
dominates over the other two channels. Note that, for
$(\lambda/k\delta) \gg 1$, the effective mass of $Z$, $\omega$, is
much larger than the weak scale and the decay into a pair of the
gravitinos is kinematically allowed.

First let us consider the case of $(\lambda/k\delta) \gg 1$, in which
the gravitino production is the main decay mode.  The total decay rate
of the $Z$ field is given by $\Gamma_Z^{\rm (total)} \simeq \Gamma_Z(Z
\rightarrow 2 \psi_{3/2})$.  Then, after the decay of $Z$, the
universe is dominated by the gravitinos with a small amount of the
entropy produced by the decay into the gluons and the gluinos.  The
gravitino-to-entorpy ratio is
\bea Y_{3/2}^{(Z)} &\simeq & \frac{2B_{3/2}}{(1-B_{3/2})^{3/4}}
 \frac{3}{4 \omega} \lrfp{\pi^2 g_*}{10}{-\frac{1}{4}} \sqrt{\Gamma^{\rm (total)}_Z } ,\non\\
&\gg&9 \times 10^{-8} \lrfp{g_*}{10.75}{-\frac{1}{4}}
\lrfp{\alpha_3}{0.1}{-\frac{3}{2}} \lrfp{m_3}{\rm TeV}{\frac{3}{2}}
\lrfp{m_{3/2}}{{\rm TeV}}{-1},
\label{eq:Y32}
\eea
where $g_*$ counts the relativistic degrees of freedom, and
$B_{3/2}\simeq 1$ denotes the branching ratio of the gravitino
production.  Thus the gravitino abundance is too large to be
compatible with the constraints from BBN which range from
$O(10^{-16})$ to $O(10^{-14})$ for $m_{3/2} = 100{\rm\,GeV} -
10{\rm\,TeV}$~\cite{Kawasaki:1999na,Hannestad:2004px} (see also
(\ref{eq:unstable-Y2})).

On the other hand, the gluino production dominates for $0.1 \lesssim
 (\lambda/k\delta) \lesssim O(1)$.  Since the decay temperature is
 rather low, the resultant lightest SUSY particles (LSP) will easily
 overclose the universe.  In order to avoid the overproduction of the
 LSP, the effective mass $\omega$ should be smaller than $2 m_3$,
 i.e., $(\lambda/k\delta) \lesssim 0.1$. Using (\ref{eq:gammag}) as
 the total decay rate, the decay temperature becomes
\bea 
T_d & \equiv & \lrfp{\pi^2 g_*}{10}{-\frac{1}{4}} \sqrt{\Gamma^{\rm (total)}_Z },\non\\
&\simeq & 0.2 {\rm\,MeV} \lrfp{g_*}{10.75}{-\frac{1}{4}}
\lrfp{\lambda}{\delta k}{\frac{3}{2}}
\lrfp{\alpha_3}{0.1}{-\frac{3}{2}} \lrfp{m_3}{\rm TeV}{\frac{5}{2}}
\lrfp{m_{3/2}}{{\rm TeV}}{-1}, \non\\
&\lesssim&5 {\rm\, keV} \lrfp{g_*}{10.75}{-\frac{1}{4}}
 \lrfp{m_3}{\rm TeV}{\frac{5}{2}}
\lrfp{m_{3/2}}{{\rm TeV}}{-1}. 
\label{eq:td}
\eea
Since the gaugino mass is proportional to the gravitino mass, the
decay temperature increases as the gravitino mass becomes larger.
However, even for $m_{3/2} \simeq m_3 = O(10)$TeV, it is still smaller
by several orders of magnitude than the lower bound on the decay
temperature by BBN~\cite{Kawasaki:1999na,Hannestad:2004px}, the cosmic
microwave background and the large scale
structure~\cite{Ichikawa:2006vm}.  Thus we conclude that once $Z$
field deviates from the origin during inflation by $\Delta Z \simeq
\delta M$, the Polonyi problem associated with the $Z$ field spoils
the success of the standard cosmology.

Now let us turn to a low-scale inflation scenario, satisfying $H_{\rm
inf} <m_{Z}$.  If the initial position of $Z$ is beyond $Z_*$, it will
settle down at $|Z| \sim |\delta M|$ during inflation, and the Polonyi
problem jeopardizes the scenario as before. In the following we assume
that this is not the case. The shift of the minimum is then given by
\begin{equation}
       |\Delta Z|\simeq \frac{H_{\rm inf}^2}{m_Z^2}|\delta M|.
       \label{eq:deltaZ-lowinf}
\end{equation}
We require $|\Delta Z| < |Z_*|$ to avoid the above Polonyi problem, which 
leads to a constraint
\begin{equation}
       H_{\rm inf} \;\lesssim\; 2 \times 10^6 {\rm\,GeV}
                        \frac{\lambda^{\frac{7}{4}}}{\sqrt{\delta k}}
                        \lrfp{\alpha_3}{0.1}{-\frac{1}{2}}
                        \lrfp{m_3}{\rm TeV}{\frac{1}{2}}
                        \lrfp{m_{3/2}}{\rm TeV}{\frac{1}{4}}.
				\label{eq:Hubble}
\end{equation}
After inflation, the $Z$ starts to oscillate with an initial amplitude
$\Delta Z$ given by (\ref{eq:deltaZ-lowinf}), and soon decays into a
pair of the gravitinos, since the rate is enhanced 
especially when $m_Z$ is much larger than $m_{3/2}$ (see the equation
above (\ref{eq:gammaz})). The abundance of the gravitinos produced by
the $Z$ decay is
\begin{eqnarray}
\label{eq:Y32-2-1}
Y_{3/2}^{(Z)} &\simeq& 1 \times 10^{-12} 
                        \frac{\delta^2 k^2}{\lambda^{\frac{15}{2}}} 
                        \lrfp{\alpha_3}{0.1}{2} 
                        \lrfp{m_{3}}{\rm TeV}{-2} 
                        \lrfp{H_{\rm inf}}{10^6{\rm\,GeV}}{2}
                        \lrf{T_R}{10^6{\rm\,GeV}} 
                        \lrfp{m_{3/2}}{\rm TeV}{\frac{1}{2}}, \\
			&\lesssim& 6 \times 10^{-12} \frac{\delta k}{\lambda^{4}}
                        \lrf{\alpha_3}{0.1} \lrfp{m_{3}}{\rm TeV}{-1}
                        \lrf{T_R}{10^6{\rm\,GeV}} \lrf{m_{3/2}}{\rm TeV},
                        \label{eq:Y32-2}
\end{eqnarray}
where $T_R$ denotes the reheating temperature of the inflaton, and we
have used (\ref{eq:Hubble}) for the last inequality. We find that the
gravitino abundance is much smaller than the previous case
(\ref{eq:Y32}), and it can be compatible with the BBN bounds.  

To sum up, the Polonyi problem associated with the $Z$ field  excludes
high-scale inflation models, and only low-scale inflation models
satisfying (\ref{eq:Hubble}) may be able to circumvent the problem.

\subsection{Gravitino production from Inflaton decay}

The gravitino production from the inflaton decay is a quite generic
phenomenon. In fact, for a large-scale inflation model, the inflaton
directly decays into the SUSY breaking
sector~\cite{Endo:2006qk,Endo:2007ih}, producing the gravitinos, while
the gravitino pair production becomes effective for a low-scale
inflation model.

Let us first consider inflation models with an inflaton mass $m_\phi$
larger than the dynamical scale $4 \pi \Lambda$, which is typically
the case for large-scale inflation like a hybrid inflation
model~\cite{Copeland:1994vg}.  The inflaton then decays into the SUSY
breaking sector through the following processes. As pointed out in
Ref.~\cite{Endo:2006qk}, the inflaton decays via the Yukawa coupling
(\ref{eq:zqq}), producing the scalar and fermionic components of $Z$
and the hidden (s)quarks $Q^i$. Note that the fermionic components of
$Z$ is the goldstino which will be eaten by the gravitino, and that
the scalar $Z$ dominantly decays into a pair of the gravitinos. In
addition, the inflaton decays into the $SP(1)$ gauge sector via the
anomalies in SUGRA~\cite{Endo:2007ih}.  The produced gauge
bosons/gauginos form jets, producing the $SP(1)$ hadrons. In the
decays of the $SP(1)$ hadrons, the gravitinos are produced.  The
detailed analyses~\cite{Endo:2006qk,Endo:2007ih} actually show that
the gravitino production through the above processes excludes almost
entire parameter spaces for the large-scale inflation models.
Therefore, even if one circumvents the Polonyi problem discussed in
the previous section (e.g. by fine-tuning the linear term of $Z$ in
the K\"ahler potential), the high-scale inflation models still suffer
from the severe gravitino overproduction problem.

In the rest of this section, we focus on low-scale inflation models
with the inflaton mass smaller than $4 \pi \Lambda$. Then the
spontaneous decays~\cite{Endo:2006qk,Endo:2007ih} do not occur, since
the SUSY breaking fields typically have a mass of the dynamical
scale. Instead, we need to take account of the gravitino pair
production from the inflaton. The gravitino pair production occurs
even when the inflaton has the minimal K\"ahler potential.  When the
K\"ahler potential is minimal for the inflaton field, the gravitino
pair production rate is~\cite{Endo:2006tf}
\beq
\Gamma_{\phi}^{(0)}(\phi \rightarrow 2 \psi_{3/2})\;\simeq\;
\frac{1}{32 \pi} \la \phi \ra^2 m_\phi^3 \times F^{(0)}
\label{eq:grav-inf-0}
\eeq
with
\beq
F^{(0)}\; \equiv\;
\left\{
\bear{cc}
\ds{ (m_Z/m_\phi)^4}
 &{\rm for~~~~}m_\phi \gg m_Z \\
\ds{1}&{\rm for~~~~}m_\phi \ll m_Z 
\eear
\right.,
\eeq
where $\la \phi \ra$ denotes the VEV of the inflaton, and the upper
index $(0)$ is to remind us that the rate is for the minimal K\"ahler
potential. Although the gravitino production is suppressed if $m_\phi
\gg m_Z$, one cannot expect too large hierarchy between $m_\phi$ and
$m_Z$, because we are considering low-scale inflation models with
$m_\phi < 4\pi\Lambda$.

Now we consider non-renormalizable couplings of the inflaton with $Z$.
Actually, those couplings that induce the mixings between $\phi$ and
$Z$ in SUGRA enhance the gravitino production.  The relevant mixings
arise from the following interactions:
\beq
K \;=\; \beta_1 |\phi|^2 Z 
		+ \frac{\beta_2}{2} |\phi|^2 Z Z + {\rm h.c.},
\eeq
where $\beta_{1(2)}$ are numerical coefficients. The presence of those
interactions is rather generic, since even though we have assumed that
such $U(1)$-breaking operators are absent at the cutoff scale, these
are radiatively induced during the evolution running down to lower
energy scale through the U(1)-breaking operator in
Eq.(\ref{eq:messenger}).  At the inflaton mass scale $m_\phi \ll M$,
$\beta_{1(2)}$ are estimated as
\bea
\beta_1 &\sim& \frac{5 \zeta}{16 \pi^2} k M^* \ln {M_P^2 \over |M|^2},\\
\beta_2 &\sim& \frac{5 \zeta}{16 \pi^2} \frac{(k M^*)^2}{|M|^2},
\eea
where we have introduced a non-renormalizable interaction,
\beq
K\;=\; \zeta |\phi|^2 |\Psi|^2,
\eeq
with a numerical coefficient $\zeta$ of order unity.  The decay rate
into a pair of the gravitinos is~\cite{Endo:2006tf}
\bea
\Gamma_{\phi}^{(1)}(\phi \rightarrow 2 \psi_{3/2}) &\simeq&
\frac{1}{32 \pi} \la \phi \ra^2 m_\phi^3 \times F^{(1)}
\label{eq:grav-inf-1}
\eea
with
\beq
F^{(1)} \;\equiv\;
\left\{
\bear{cc}
\ds{\frac{|\beta_1|^2}{3} \frac{m_Z^4}{m_{3/2}^2 m_\phi^2} + |\beta_2|^2}&
{\rm for~~~~}m_\phi \gg m_Z \\
&\\
\ds{\frac{|\beta_1|^2}{3} \frac{m_\phi^2}{m_{3/2}^2} + |\beta_2|^2 
\lrfp{m_\phi}{m_Z}{4}}&{\rm for~~~~}m_\phi \ll m_Z 
\eear
\right.,
\eeq
where we have neglected interference terms for simplicity.

Combining (\ref{eq:grav-inf-0}) and (\ref{eq:grav-inf-1}), the
gravitino abundance becomes
\bea
Y^{(\phi)}_{3/2} &\simeq&
2 \frac{\Gamma_\phi(\phi \rightarrow 2 \psi_{3/2})}{\Gamma^{\rm (total)}_\phi} 
\frac{3 T_R}{4 m_\phi},\non\\
&\simeq&7 \times 10^{-15} F \lrfp{g_*}{200}{-\frac{1}{2}} 
\lrfp{\la \phi \ra}{10^{15}{\rm GeV}}{2}
\lrfp{m_\phi}{10^{10}{\rm GeV}}{2} \lrfp{T_R}{10^6{\rm GeV}}{-1}, 
\label{eq:Ytotal}
\eea
where $\Gamma^{\rm (total)}_\phi$ is the total decay rate of the
inflaton, and is related to the reheating temperature $T_R$ as
\beq
\Gamma^{\rm (total)}_\phi = \lrfp{\pi^2 g_*}{10}{\frac{1}{2}} T_R^2.
\eeq
We have also defined $F \equiv F^{(0)}+F^{(1)}$ by ignoring the interferences. 

Let us compare the result (\ref{eq:Ytotal}) to the gravitino
production from the $Z$ field associated with the Polonyi problem. We
notice that the dependence of $Y^{(\phi)}_{3/2}$ on the reheating
temperature is different from that of $Y^{(Z)}_{3/2}$ in
(\ref{eq:Y32-2-1}). Importantly, the gravitino overproduction problem
cannot be solved simply by reducing the reheating temperature, since
$Y^{(\phi)}_{3/2}$ is inversely proportional to $T_R$.  This makes it
nontrivial whether there exist cosmologically allowed parameter
regions.

\subsection{Example}

Let us consider a new inflation model~\cite{Izawa:1996dv} as an
example of the low-scale inflation models. The K\"ahler potential and
superpotential of the inflaton sector are written as
\bea
K(\phi,\phi^\dag) &=& |\phi|^2 + \frac{\kappa}{4 } |\phi|^4,\non\\
W(\phi) &=&  v^2\phi -  \frac{g}{n+1}\, \phi^{n+1}.
\label{IY}
\eea
where the observed density fluctuations are explained for $v \simeq
4\times 10^{-7} \, (0.1/g)^{1/2}$ and $\kappa \lesssim 0.03$ in the
case of $n=4$. We assume $n=4$ in the following, since the Hubble
parameter during inflation likely exceeds the bound (\ref{eq:Hubble})
for $n > 4$~\cite{Ibe:2006fs}.  After inflation, the inflaton $\phi$
takes the expectation value $\la\phi\ra \simeq (v^2/g)^{1/n}$.  The
inflaton mass is given by $m_{\phi} \simeq n v^2/\la\phi\ra$, and the
gravitino mass is related to $v$ as $m_{3/2} \simeq n v^2 \la\phi\ra
/(n+1) $, since the inflaton VEV induces the spontaneous breaking of
the $R$-symmetry, namely a nonzero $\langle W \rangle$.  Precisely
speaking, $v$ has a weak dependence on $T_R$ via an e-folding number
and on $\kappa$, and $\la\phi\ra$ depends on these parameters as
well. Therefore one can express the inflation scale $v$ and the
coupling $g$ as functions of $m_{3/2}$, $T_R$ and $\kappa$, once the
WMAP normalization of the density fluctuations~\cite{Spergel:2006hy}
is applied. In the following numerical analyses, we take into account
these corrections.

In the numerical analyses, we estimate the gravitino abundance.  In
addition to the non-thermally produced gravitinos,  (\ref{eq:Y32-2-1})
and (\ref{eq:Ytotal}), we also include the contribution from
the thermally produced gravitinos:~\cite{Bolz:2000fu}\footnote{ 
The gluon in the hot
plasma might decays into the gravitino due to the thermal
corrections~\cite{Rychkov:2007uq}.  This, however, changes the
gravitino production rate only by a factor.  
}
\begin{eqnarray}
    \label{eq:Yx-new}
    Y_{3/2} &\simeq& 
    1.9 \times 10^{-12}
    \left( \frac{T_{\rm R}}{10^{10}\ {\rm GeV}} \right)
    \nonumber \\ 
    & \times & 
    \left[ 1 
        + 0.045 \ln \left( \frac{T_{\rm R}}{10^{10}\ {\rm GeV}} 
        \right) \right]
    \left[ 1 
        - 0.028 \ln \left( \frac{T_{\rm R}}{10^{10}\ {\rm GeV}} ,
        \right) \right].
\end{eqnarray}
The gravitino abundance is severely 
constrained by BBN as~\cite{Kawasaki:2004yh}
\begin{eqnarray}
\label{eq:unstable-Y1}
   Y_{3/2}  & ~ \lesssim & \left\{\begin{array}{lcl}
   ~1\times 10^{-16} - 6\times 10^{-16}
   &{\rm for}    &  m_{3/2} \simeq 0.1 - 0.2~{\rm TeV} \\[0.8em]
   ~4\times 10^{-17} - 6\times 10^{-16}
   &{\rm for}    &  m_{3/2} \simeq 0.2 - 2~{\rm TeV} \\[0.8em]
   ~ 7 \times 10^{-17} - 2\times 10^{-14} 
   &{\rm for}    & m_{3/2} \simeq 2 - 10~{\rm TeV}
\end{array}\right.
\label{eq:unstable-Y2}
\end{eqnarray}
for the unstable gravitino with a hadronic branching ratio $B_h \simeq 1$.

\begin{figure}[t!]
\begin{center}
\includegraphics[width=12cm]{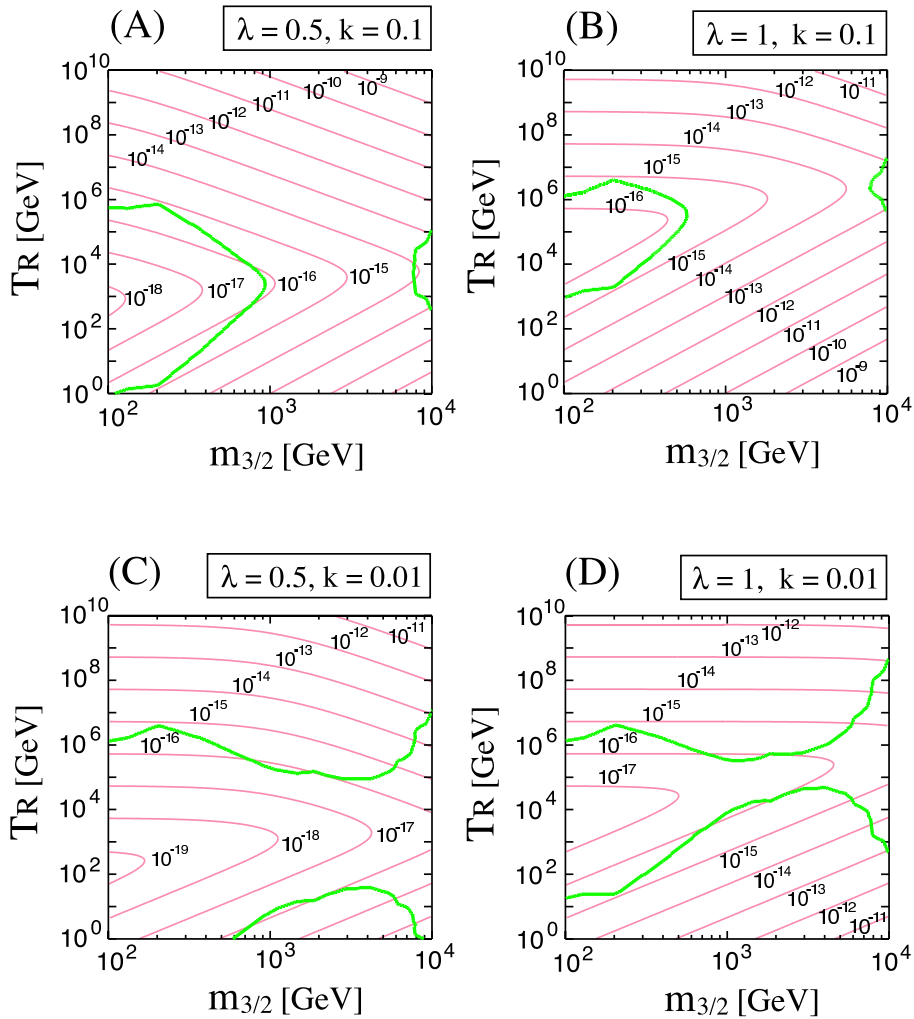}
\caption{Contours of the gravitino abundance $Y_{3/2}$ and the BBN
constraints on the $(m_{3/2}, \,T_R)$ plane, in the case of the new 
inflation model. The regions surrounded 
by the thick solid (green) lines are consistent with the BBN constraints 
(\ref{eq:unstable-Y2}). We have set the parameters as: $\delta = 0.1$ 
and $\kappa = 10^{-2}$,
varying  $(\lambda, k)$ as $(\lambda, k) = (0.5, 0.1), (1, 0.1), (0.5, 0.01)$ 
and $(1, 0.01)$ labeled by (A), (B), (C) and (D), respectively.
}
\label{fig:contour}
\end{center}
\end{figure}

We show contours of the gravitino abundance $Y_{3/2}$ for the new
inflation model (thin solid (red) lines) in Fig.~\ref{fig:contour},
for several sets of $(\lambda, k) = (0.5, 0.1), (1, 0.1), (0.5, 0.01)$
and $(1, 0.01)$ labeled by (A), (B), (C) and (D), respectively.  We
also show the parameter space consistent with the BBN bounds
(\ref{eq:unstable-Y2}), which is enclosed by the thick solid (green)
lines. We have chosen the other parameters of the SUSY breaking sector
as $\delta = 0.1$ and $\zeta = 1$. We  have taken $\eta = 1$,
which should be in principle determined by the strong dynamics, and
$m_3 = 1{\rm TeV}$ as a reference value. The parameters of the new
inflation model, $\kappa$ and $n$, are chosen as $\kappa = 10^{-2}$
and $n = 4$, while $g$ is not an independent parameter and is
determined by the other parameters. The reheating temperature $T_R$ is
regarded as a free parameter, by assuming proper couplings of the
inflaton with the SM sector~\cite{Ibe:2006fs}, though the spontaneous
decays via the SUGRA effects~\cite{Endo:2006qk} provides the lowest
reheating temperature as $T_R \gtrsim O(1)$GeV.

 From Fig.~\ref{fig:contour}, one can see that the gravitino abundance
becomes larger for the heavier gravitino.  It also tends to increase
when $T_R$ is both raised and lowered. These behaviors are mainly due
to the non-thermal productions of the gravitinos, (\ref{eq:Y32-2-1})
and (\ref{eq:Ytotal}), and it can be understood as follows. For larger
$m_{3/2}$ and $T_R$, the gravitino production from the SUSY breaking
field becomes important, while the gravitino from an inflaton
dominates for smaller $T_R$.  Here it should be noted that the
inflaton mass is positively correlated with the gravitino mass in the
model (\ref{IY}).

Since the gravitino abundance is sensitive to the model parameters,
the allowed regions change for a different set of $(\lambda, k)$.
When $\lambda$ increases, the gravitino production from the SUSY
breaking field (\ref{eq:Y32-2-1}) is suppressed, while more gravitinos
are produced by the inflaton decay (see (\ref{eq:Ytotal})).  Thus the
allowed region shifts upwards as can be seen by comparing the panel
(B) with (A) (or (D) with (C)) in Fig.~\ref{fig:contour}. On the other
hand, as $k$ becomes smaller, the abundances of the gravitino produced
both from the $Z$ and the inflaton decrease (see (\ref{eq:Y32-2-1})
and (\ref{eq:Ytotal})).  Comparing the panel (C) with (A) (or (D) with
(B)) in Fig.~\ref{fig:contour}, we find much broader allowed region
for smaller $k$.

We  notice that for smaller $m_{3/2}$ and relatively large $T_R$,
the contours of the gravitino abundance tends to become independent of
$m_{3/2}$, especially when the non-thermal gravitino production is
suppressed.  This means that the abundance is dominated by the thermal
production, and therefore it is determined solely by $T_R$ there.

One may be interested in the regions of higher reheating temperature
such as $T_R \gtrsim 10^6$ GeV, where non-thermal
leptogenesis~\cite{Fukugita:1986hr,Asaka:1999yd,Lazarides:1993sn} may
be able to explain the baryon asymmetry of the universe. For the panel
(C) in Fig.~\ref{fig:contour}, we find allowed regions with $T_R
\gtrsim 10^6\,$GeV for $m_{3/2} \gtrsim 7\,$TeV and $m_{3/2} = 100 -
400\,$GeV, and similarly, for $m_{3/2} \gtrsim 4\,$TeV and $m_{3/2} =
100 - 500\,$GeV in the panel (D).  Notice that the thermal production
of the gravitino (\ref{eq:unstable-Y2}) imposes constraints on the
reheating temperature as $T_R \lesssim {\rm a~few} \times 10^6\,$GeV
for $m_{3/2} = 100 - 500\,$GeV, while it becomes significantly relaxed
for $m_{3/2} \gtrsim 4\,$TeV.  Therefore, in the panel (D), the upper
bound on the reheating temperature is almost the same as that from the
thermal gravitino production. 

\begin{figure}[t!]
\begin{center}
\includegraphics[width=12cm]{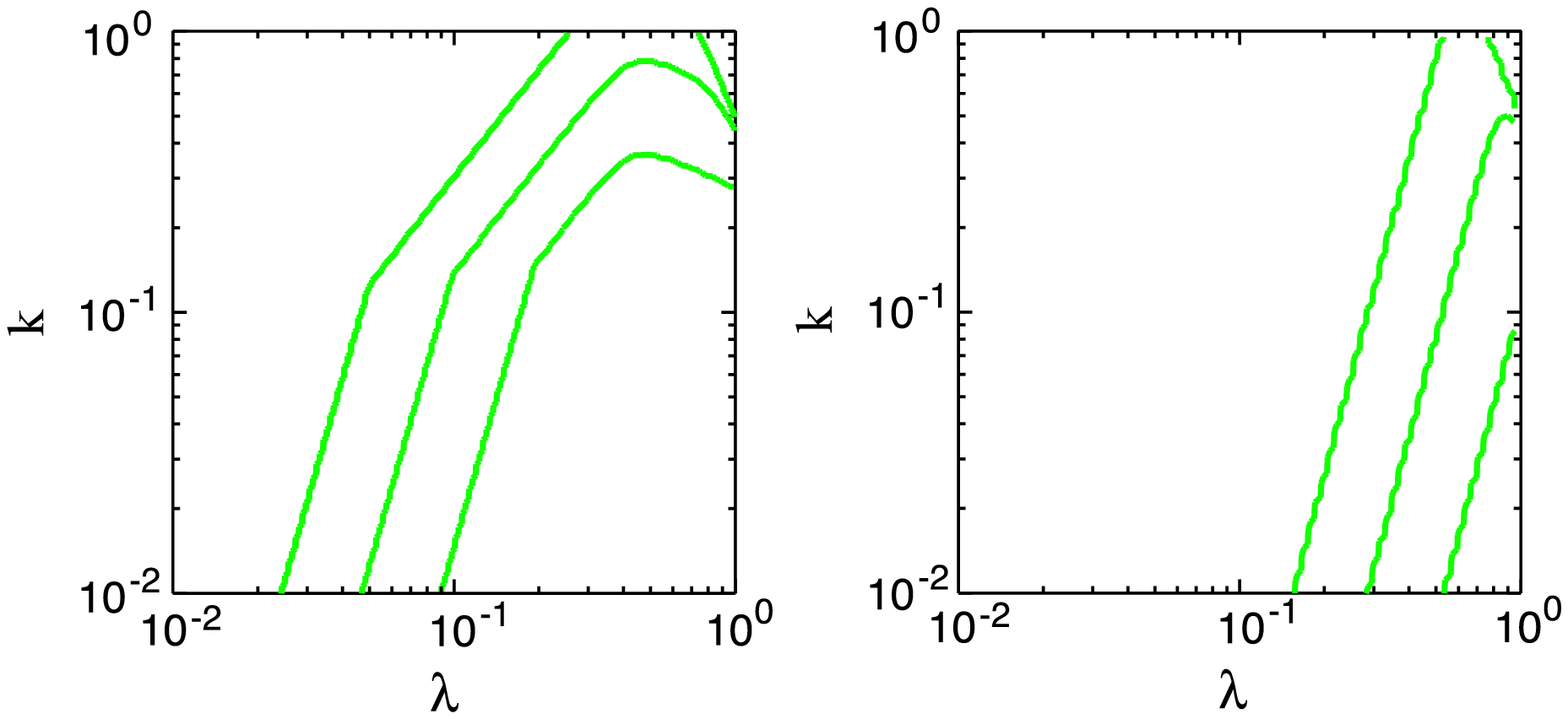}
\caption{BBN constraints on  $\lambda$ and $k$ for
$\delta = 0.01,0.1$ and $1$ (from left to right). The regions
below the lines are allowed, in the sense that
there exists a set of $(m_{3/2}, T_R)$ consistent with the BBN results 
in the range, $m_{3/2} \in [10^2,10^4]$GeV and $T_R \in [1,10^{10}]$GeV 
(left panel); $m_{3/2} \in [10^2,10^4]$GeV and $T_R \in [10^6,10^{10}]$GeV (right panel).
}
\label{fig:lambda-k}
\end{center}
\end{figure}

We have varied $\lambda$, $k$ and $\delta$ to see how the constraints
depend on the parameters in the SUSY breaking sector.  In
Fig.~\ref{fig:lambda-k}, we have examined whether or not the allowed
region exits for $m_{3/2} = 100{\rm\, GeV} -10{\rm\, TeV}$ and $T_R =
1{\rm\, GeV} - 10^{10}{\rm\, GeV}$ (left panel), and for $m_{3/2} =
100{\rm\, GeV} -10{\rm\, TeV}$ and $T_R = 10^6{\rm\, GeV} -
10^{10}{\rm\, GeV}$ (right panel). One can see that $\lambda$ is
bounded below for fixed $k$, since too small $m_Z$ makes the Polonyi
problem worse (see (\ref{eq:Y32-2-1})). Although not explicitly shown
in the figure, $\lambda$ cannot be too large since more gravitinos are
produced from the inflaton.  For smaller $k$ and $\delta$, it becomes
easier to satisfy the BBN bounds, though too small $k$ upsets the
stability of the SUSY breaking vacuum.

As a comment, although we have taken account of the LSP production
from the gravitinos, it does not give any meaningful
constraints on the parameters in which we are interested~\footnote{
Here we have neglected the thermal production in the estimation of the
LSP abundance. It is highly dependent on the mass spectrum of the
visible sector.}. Indeed, the abundance of the LSP produced from
the gravitino is always negligible for the LSP mass $\lesssim O(1)$\,TeV,
as long as we require the BBN constraints on $Y_{3/2}$ 
(\ref{eq:unstable-Y2}) to be satisfied.

So far, we have considered the new inflation model (\ref{IY}), in
which the inflaton mass and the gravitino mass are correlated.  The
relation does not hold if we consider a two-field new
inflation model~\cite{Asaka:1999jb}. For instance, even for larger
gravitino mass, we can take the inflaton mass lower than the model
(\ref{IY}).  Then the gravitino production from both the Polonyi field
and the inflaton can be suppressed.

\section{Conclusions}
\label{sec:4}

The gravity mediation provides a simple way to mediate the SUSY
breaking to the visible sector, and so, it has been one of the main
target of research. However, recent observations on the gravitino
overproduction from an inflaton and the Polonyi problem revealed that
the gravity mediation is faced with the cosmological embarrassment,
which drives the scenario into a corner, hinting that some improvement
is needed.  In this paper, we have proposed the retrofitted gravity
mediation model to circumvent the cosmological problems.  We have
introduced an approximate $U(1)$ symmetry under which the SUSY
breaking field is charged to alleviate the gravitino overproduction
from an inflaton and a supersymmetry breaking field. Indeed, we have
found such regions for a low-scale inflation model that all the
superparticles, especially, the gauginos as well as the gravitino,
have a mass around the weak scale and the cosmological bounds on the
gravitino abundance are satisfied. 

Specifically, there are allowed regions that the reheating temperature is 
larger than $10^6\,$GeV for $m_{3/2} \gtrsim 4$TeV and $m_{3/2} = 100 - 500$GeV.  These regions
are attractive since non-thermal leptogenesis may be able to explain
the baryon asymmetry of the universe.  For the gravitino mass heavier than
$4$TeV, the squark and sleptons acquire large masses of $O(m_{3/2})$,
and therefore the problems of large flavor changing neutral currents
and CP violation become mild. Further, if the gaugino (and/or higgsino) masses
are so light as $O(100)$\,GeV, the particle spectrum resembles that in the focus-point 
region~\cite{Feng:1999mn}, and the lightest neutralino may be able to
explain the current  DM abundance~\cite{Ibe:2005jf}.

Finally let us comment on the possible extension of the model.  From
the numerical analyses, we have seen that smaller $k$ and $\delta$ are
cosmologically favored since the gravitino overproduction problem gets
greatly relaxed. Such a suppression of $k$ and $\delta$ can be
realized easily by imposing the $Z_2$ symmetry.  In fact, if $Z$ and
$QQ$ are odd under $Z_2$, the Yukawa coupling $k$ and the mass $M'$
should break the $Z_2$ symmetry. Thus $k$ and $\delta = M'/M$ become
naturally small. However it is noticed that $k$ cannot be too small,
since the stability of the SUSY breaking vacuum would be upset due to
the tachyonic messengers and/or the decay via the tunneling to the
SUSY preserving vacuum.

\section*{Acknowledgment}
We thank M.~Ibe and S.~Shirai for useful discussions.

\end{document}